\documentstyle[12pt,epsfig]{article}

\newlength{\dinwidth}
\newlength{\dinmargin}
\begin{document}
\titlepage
\begin{flushright}
DTP/96/48  \\
June 1996 \\
\end{flushright}

\begin{center}
\vspace*{2cm}
{\Large \bf Observable jets from the BFKL chain} \\
\vspace*{1cm}
J.\ Kwiecinski$^{a,b}$, C.\ A.\ M.\ Lewis$^{a,b}$ and A.\ D.\
Martin$^b$ \\
\end{center}
\vspace*{0.5cm}

\begin{center}
$^a$ Department of Theoretical Physics, \\
H.\ Niewodniczanski Inst.\ of Nuclear Physics, \\
ul.\ Radzikowskiego 152, \\
31-342 Krakow, Poland. \\

\vspace*{0.5cm}
$^b$ Department of Physics, \\
University of Durham, \\
Durham, DH1 3LE, UK.
\end{center}

\vspace*{2cm}

\begin{abstract}
We derive a modified form of the BFKL equation which enables the
structure of the gluon emissions to be studied in small $x$ deep
inelastic scattering.  The equation incorporates the resummation
of the virtual and unresolved real gluon emissions.  We solve the
equation to calculate the number of small $x$ deep-inelastic
events containing 0, 1, 2 \ldots resolved gluon jets, that is
jets with transverse momenta $q_T > \mu$.  We study the jet
decomposition for different choices of the jet resolution
parameter $\mu$.
\end{abstract}

\newpage
\noindent {\large \bf 1.  Introduction} \\

The advent of the HERA electron-proton collider has opened up the
possibility of testing QCD in the new and hitherto unexplored
small $x$ regime.  The HERA measurements of the proton structure
function $F_2 (x, Q^2)$ show a striking rise with decreasing $x$,
which with the latest data is now known with considerable
precision \cite{h1,zeus}.  On the other hand from the theoretical
point of view we know for sufficiently small $x$, such that
$\alpha_S \ln 1/x \sim 1$, that it is necessary to resum the
$(\alpha_S \ln 1/x)^n$ contributions in order to obtain reliable
perturbative QCD predictions.  At leading order this is
accomplished by the Balitzkij, Fadin, Kuraev, Lipatov (BFKL)
equation \cite{bfkl}.  This equation effectively corresponds to
the sum of gluon ladder diagrams of the type shown in Fig.\ 1 in
which the transverse momenta $q_T$ are unordered along the chain.
This should be contrasted with DGLAP evolution where, in the
leading $\ln Q^2$ approximation, the transverse momenta are
strongly ordered from the hadronic to the hard scale $Q^2$ which
in deep-inelastic lepton scattering is provided by the virtuality
of the photon, namely
\begin{equation}
Q^2 \; \gg \; k_T^2 \; \gg \; k_{nT}^2 \; \gg \; \ldots
\label{eq:a1}
\end{equation}
Both BFKL and DGLAP evolution lead to an increase of the
deep-inelastic scattering structure functions with decreasing
$x$.  In fact it is possible to obtain a satisfactory description
of the rise of the structure function measured in the HERA small
$x$ regime
using both approaches \cite{MRS,GRV,CCFMF2}.  The inclusive nature 
of the structure
function $F_2$ makes it extremely difficult, even with the
precise
HERA data, to use the observed $x$ behaviour to reveal the
underlying dynamics at small $x$.  This is not surprising.  The
leading behaviour obtained from BFKL is an $x^{- \lambda}$
growth, whereas for DGLAP we anticipate an increase of the double
logarithmic form $\exp \left (A [\ln (t/t_0) \: \ln
(1/x)]^{\frac{1}{2}} \right )$ where $t = \ln (Q^2/\Lambda^2)$. 
However, these are asymptotic predictions.  For instance
subleading $\ln 1/x$ effects will weaken the BFKL growth in the
HERA regime \cite{KMSGLU,LUND}.  Moreover the DGLAP behaviour is 
dependent on the
choice of a non-perturbative input form at some scale $Q^2 =
Q_i^2$.  It has been realized however, that the intimate relation
between the increase of the cross sections with decreasing $x$
and the absence of transverse momentum ordering, which is the
basic
property of the BFKL dynamics, should reflect itself in the
properties of the final states in deep-inelastic lepton
scattering.  Indeed several dedicated measurements have been
proposed
and are being experimentally studied at HERA (see, for example,
the reviews in ref.\ \cite{fs}).

The purpose of this paper is to study the detailed properties of
the partonic final state produced by the gluon emissions along
the BFKL
chain.  In this way we will gain insight into the BFKL equation,
as well as detailing observables with which to probe the
underlying small
$x$ dynamics.  In particular we calculate the decomposition of
the
(total) deep-inelastic cross section into components $\sigma_n
(\mu)$ which correspond to the production of a fixed number $n$
of gluon jets each with transverse momentum $q_T > \mu$.  That is
we study the possible jet configurations in the central region
between the
current jet and the proton remnants.  An interesting feature of
BFKL dynamics is the possibility of producing jets even for $\mu
> Q$.  One of our aims is to quantify the yield of such jet
configurations.

In the BFKL equation there is a delicate cancellation between the
real gluon emissions and the virtual contributions.  Clearly the
cancellation is affected by the resolution $q_T > \mu$ that we
impose.  In particular we must ensure that the appropriate
cancellation between the virtual contributions and the \lq\lq
unresolved" real gluon emissions with $q_T < \mu$ is maintained
throughout the calculation.  We must therefore first derive a
modified form of the BFKL equation which will enable us to
quantify the number of energetic {\it resolved} jets produced
along the gluon chain, but in which the {\it virtual} and {\it
unresolved} contributions are treated on an equal footing and
are resummed.  This is the subject of section 2.  In section 3 we
give an analytic solution for the resummation at low jet 
resolution, whereas in section 4 we consider more realistic values 
of the resolution $\mu$ and solve the modified BFKL equation by 
iteration to illustrate the jet decomposition of the BFKL gluon.  
At this stage it is still a theoretical study.  In section 5 we use 
the BFKL gluon and the $k_T$-factorization theorem \cite{ktfac} to 
predict the jet decomposition of the observable structure function 
$F_2$ and the deep-inelastic cross section.  Section 6 contains our 
conclusions. \\

\bigskip
\noindent {\large \bf 2.  The BFKL equation incorporating jet
resolution $q_T > \mu$}

In the small $x$ regime the dominant parton is the gluon.  Since
we no longer have strong-ordering in transverse momenta along the
gluon chain in Fig.\ 1 we must work in terms of the gluon
distribution $f (x, k_T^2)$ unintegrated over its transverse
momentum $k_T$.  The relation of unintegrated distribution $f$ to
the conventional gluon distribution is
\begin{equation}
g (x, Q^2) \; = \; \int^{Q^2} \: \frac{dk_T^2}{k_T^2} \: f (x,
k_T^2).
\label{eq:a2}
\end{equation}
The unintegrated density $f$ satisfies the BFKL equation which
effectively sums up the leading $\alpha_S \ln 1/x$ contributions.
In integral form it may be written \cite{CIAF,CCFM}
\begin{equation}
f (y, k_T^2) \; = \; f^{(0)} (y, k_T^2) \: + \:
\overline{\alpha}_S \: \int_0^y \: dy^\prime \: \int \: \frac{d^2
q_T}{\pi q_T^2} \: \left [ \frac{k_T^2}{k_T^{\prime 2}} \: f
(y^\prime, k_T^{\prime 2}) \: - \: f (y^\prime, k_T^2) \: \Theta
(k_T^2 - q_T^2) \right ],
\label{eq:a3}
\end{equation}
with $\overline{\alpha}_S \equiv 3 \alpha_S/\pi$.  We have chosen
to use the rapidity variable $y = \ln 1/x$ instead of $x$,  and
so the integral in (\ref{eq:a3}) has come from the replacement
$$
\int_x^1 \: \frac{dx^\prime}{x^\prime} \; \rightarrow \; \int_0^y
\: dy^\prime .
$$
For convenience we have also introduced
\begin{equation}
k_T^{\prime 2} \; \equiv \; \left | \mbox{\boldmath $q$}_T \: +
\: \mbox{\boldmath $k$}_T \right |^2 .
\label{eq:a4}
\end{equation}
Note that the dependence on $k_T^{\prime 2}$ makes the angular
integration in $d^2 q_T$ non-trivial.  The inhomogeneous
contribution $f^{(0)}$ in (\ref{eq:a3}) corresponds to the \lq\lq
no-rung" contribution of Fig.\ 1.  It is the driving term of the
equation and has to be input.  We implicitly include under the
$d^2 q_T$ integral in (\ref{eq:a3}) the product of theta
functions
\begin{equation}
\Theta (Q_f^2 \: - \: k_T^{\prime 2}) \: \Theta (k_T^{\prime 2}
\: - \: Q_0^2)
\label{eq:a5}
\end{equation}
so that the emitted gluon is constrained to the domain $Q_0^2 <
k_T^{\prime 2} < Q_f^2$.  In the numerical predictions shown
below we take $Q_0^2 = 1$ GeV$^2$ and $Q_f^2 = 10^4$ GeV$^2$.

Jet structure is embodied in the BFKL equation via real gluon
emission from the gluon chain prior to its interaction with the
photon probe (which takes place through the usual fusion
subprocess
$\gamma g \rightarrow q\overline{q}$).  An observed jet is
defined by a resolution parameter $\mu$ which specifies the
minimum transverse momentum that must be carried by the emitted
gluon for it to be detected.  For realistic observed jets in the 
experiments at HERA, the lowest reasonable choice for the
resolution cut-off parameter $\mu$ appears to be about $\mu =
3.5$ GeV.  However, we also present results for $\mu = 6$ GeV
and, so as to gain theoretical insight, for the low values of $\mu
= 1$ and $2$ GeV.

If an emitted gluon has transverse momentum $q_T < \mu$ then the
radiation is said to be unresolved.  The unresolved radiation
must be treated at the same level as the virtual corrections to
ensure that the singularities as $q_T^2 \rightarrow 0$ cancel in
the $q_T^2$ integration.  To do this we first rewrite the BFKL
equation (\ref{eq:a3}) in the symbolic form
\begin{equation}
f \; = \; f^{(0)} \: + \: \int_0^y \: dy^\prime \: K \otimes f
(y^\prime),
\label{eq:a6}
\end{equation}
where $\otimes$ denotes the convolution over $q_T$.  We divide
the real gluon emission contribution into resolved
and unresolved parts using the identity
\begin{equation}
\Theta (q_T^2 - \mu^2) \: + \: \Theta (\mu^2 - q_T^2) \; = \; 1,
\label{eq:a7}
\end{equation}
where the first term denotes the real resolved emission and the 
second the real unresolved emission. We then combine the unresolved 
component with the virtual
contribution \cite{JAR}.  That is
\begin{equation}
f \; = \; f^{(0)} \: + \: \int_0^y \: dy^\prime \: (K_R + K_{UV})
\: \otimes \: f (y^\prime),
\label{eq:a8}
\end{equation}
where the kernel $K_R$ for the {\it resolved} emissions with $q_T
> \mu$ is given by
\begin{equation}
K_R \: \otimes \: f(y^\prime) \; = \; \overline{\alpha}_S (k_T^2)
\: k_T^2 \: \int \: \frac{d^2 q_T}{\pi q_T^2} \: \Theta (q_T^2 -
\mu^2) \: \frac{1}{k_T^{\prime 2}} \: f (y^\prime, k_T^{\prime
2}),
\label{eq:a9}
\end{equation}
while $K_{UV}$, the combined {\it unresolved} and {\it virtual}
part of the kernel, satisfies
\begin{equation}
K_{UV} \: \otimes \: f (y^\prime) \; = \; \overline{\alpha}_S
(k_T^2) \: \int \: \frac{d^2 q_T}{\pi q_T^2} \: \left [
\frac{k_T^2}{k_T^{\prime 2}} \: f (y^\prime, k_T^{\prime 2}) \:
\Theta (\mu^2 - q_T^2) \: - \: f (y^\prime, k_T^2) \: \Theta
(k_T^2 - q_T^2) \right ],
\label{eq:a10}
\end{equation}
with $k_T^{\prime 2} \equiv |\mbox{\boldmath $q$}_T +
\mbox{\boldmath $k$}_T |^2$.  These identifications of the
kernels follow by comparing (\ref{eq:a8}) with (\ref{eq:a3}). 
The $q_T^2 \rightarrow 0$ singularity is now cancelled between
the unresolved and virtual contributions, and by working with the
combined kernel $K_{UV}$ we will ensure that the cancellation
remains intact.

We seek a BFKL equation for the real resolved emissions in which
the unresolved and virtual contributions have been resummed.  To
do this we write the BFKL equation (\ref{eq:a8}) in the
differential form
\begin{equation}
\frac{\partial f}{\partial y} \; = \; \left ( \frac{\partial
f^{(0)}}{\partial y} \; + \; K_R \: \otimes \: f \right ) \; + \;
K_{UV} \: \otimes \: f,
\label{eq:a11}
\end{equation}
and treat the expression in brackets as the inhomogeneous
contribution.  We solve the inhomogeneous equation in the
standard way.  We first find a solution to the homogeneous
equation and then we obtain the full solution via an integrating
factor.  The homogeneous version of (\ref{eq:a11}) is
\begin{equation}
\frac{\partial \Delta}{\partial y} \; = \; K_{UV} \: \otimes \:
\Delta
\label{eq:a12}
\end{equation}
with solution
\begin{equation}
\Delta (y) \; = \; \exp (y \: K_{UV}),
\label{eq:a13}
\end{equation}
and so the integrating factor is $\Delta^{-1} = \exp (-y \: 
K_{UV}).$  Hence the full solution of (\ref{eq:a11}) is
\begin{eqnarray}
f (y) & = & \int_0^y \: dy^\prime \: \Delta (y) \: \otimes \:
\Delta^{-1} (y^\prime) \: \otimes \: \left ( \frac{\partial
f^{(0)}}{\partial y^\prime} \: + \: K_R \: \otimes \: f
(y^\prime) \right ) \nonumber \\
& & \nonumber \\
& = & \int_0^y \: dy^\prime \: e^{(y - y^\prime) K_{UV}} \:
\otimes \: \left ( \frac{\partial f^{(0)}}{\partial y^\prime} \:
+ \: K_R \: \otimes \: f (y^\prime) \right ).
\label{eq:a14}
\end{eqnarray}
Thus we have derived a BFKL equation for the gluon distribution
$f$ in which the unresolved and virtual terms have been resummed
in the exponential factor.  The equation is of the form
\begin{equation}
f (y) \; = \; \hat{f}^{(0)} (y) \: + \: \int_0^y \: dy^\prime \:
\hat{K} \: \otimes \: f (y^\prime)
\label{eq:a15}
\end{equation}
where the driving term has become
\begin{equation}
\hat{f}^{(0)} (y) \; = \; \int_0^y \: dy^\prime \: e^{(y -
y^\prime) K_{UV}} \: \otimes \: \frac{\partial f^{(0)}}{\partial
y^\prime}
\label{eq:a16}
\end{equation}
and the new kernel
\begin{equation}
\hat{K} \; = \; e^{(y - y^\prime) K_{UV}} \: \otimes \: K_R.
\label{eq:a17}
\end{equation}
Recall that the original BFKL kernel, $K_R + K_{UV}$, has no $y$
(i.e.\ $x$) dependence.  However, upon the resummation of the
unresolved and virtual radiation we generate an explicit $y$
dependence.  In fact the kernel $\hat{K}$ of (\ref{eq:a15}) is a
function of only the difference $y - y^\prime$ (i.e.\ of $\ln
x^\prime/x$) and not $y$ and $y^\prime$ individually, see
(\ref{eq:a17}). \\

\bigskip
\noindent {\large \bf 3.  Analytical solution at low $\mu$}

In section 4 we numerically solve the modified BFKL equation for
$f (y, k_T^2)$ and, by iteration, determine the probability
of the emission of $n$ gluon jets with $q_T > \mu$.  However,
first it is informative to derive an approximate form of the
above equation which holds in the (theoretical) limit of small
$\mu^2/k_T^2$.  In this limit it is possible to resum the
unresolved and virtual contributions in a closed analytic form. 
The crucial observation is that for small $\mu^2/k_T^2$ we may
write
$$
k_T^{\prime 2} \; \equiv \; | \mbox{\boldmath $q$}_T \: + \:
\mbox{\boldmath $k$}_T |^2 \; \approx \; k_T^2
$$
in the integrand for the unresolved real emission term in
(\ref{eq:a10}).  Then (\ref{eq:a10}) simplifies to become
\begin{eqnarray}
K_{UV} \otimes f(y^\prime) & = & \overline{\alpha}_S (k_T^2) \: f
(y^\prime) \: \int \: \frac{dq_T^2}{q_T^2} \: \left [ \Theta
(\mu^2 - q_T^2) \: - \: \Theta (k_T^2 - q_T^2) \right ] \: + \:
{\cal O} \left (\frac{\mu^2}{k_T^2} \right ) \nonumber \\
& & \nonumber \\
& = & - \overline{\alpha}_S (k_T^2) \: \ln \left (
\frac{k_T^2}{\mu^2} \right ) \: f (y^\prime) \: + \: {\cal O}
\left ( \frac{\mu^2}{k_T^2} \right ).
\label{eq:b17}
\end{eqnarray}
Thus the homogeneous solution of the BFKL equation (\ref{eq:a11})
is
\begin{equation}
\Delta (y) \; = \; \exp (y \: K_{UV}) \; = \; \exp \biggl (
-y \: \overline{\alpha}_S (k_T^2) \: \ln (k_T^2/\mu^2) \biggr ),
\label{eq:c17}
\end{equation}
that is the resummation is given by a simple analytic form.  As a
consequence, in the small $\mu$ limit, the modified BFKL equation
(\ref{eq:a15}) becomes
\begin{equation}
f (y, k_T^2) \; = \; \hat{f}^{(0)} (y, k_T^2) \: + \:
\overline{\alpha}_S (k_T^2) \: \int_0^y \: dy^\prime \: \Delta (y
- y^\prime, k_T^2) \: \int \: \frac{d^2 q_T}{\pi q_T^2} \: \Theta
(q_T^2 - \mu^2) \: \frac{k_T^2}{k_T^{\prime 2}} \: f (y^\prime,
k_T^{\prime 2})
\label{eq:d17}
\end{equation}
where here $k_T^{\prime 2} = | \mbox{\boldmath $q$}_T +
\mbox{\boldmath $k$}_T |^2$, and the driving term is given by
\begin{equation}
\hat{f}^{(0)} (y, k_T^2) \; = \; \int_0^y \: dy^\prime \: \Delta
(y - y^\prime, k_T^2) \: \frac{\partial f^{(0)} (y^\prime,
k_T^2)}{\partial y^\prime}.
\label{eq:e17}
\end{equation}
Of course for the results presented below we do not use the low
$\mu$ approximation, although to gain insight we will compare the
full prediction of (\ref{eq:a16}) for $\hat{f}^{(0)}$ with the
approximate ${\cal O} (\mu^2/k_T^2)$ result given in
(\ref{eq:e17}).

\bigskip
\noindent {\large \bf 4.  Jet decomposition of the BFKL gluon}

The BFKL equation was expressed in form (\ref{eq:a15})
specifically so that we can decompose the unintegrated gluon 
distribution $f$ into the sum of contributions with different numbers of 
{\it resolved} gluon jets with transverse momenta $q_T > \mu$.  That
is
\begin{equation}
f (y) \; = \; \sum_{n = 0}^\infty \: f^n (y)
\label{eq:a18}
\end{equation}
where $f^n$ denotes the contribution to the unintegrated gluon
distribution $f$ arising from $n$ resolved jets in the chain,
each with $q_T > \mu$, see Fig.\ 2.  The $n$-jet contribution $f^n$ 
obviously depends on the resolution $\mu$, whereas the sum $f$ does not.  
Using (\ref{eq:a15}) we have
\begin{equation}
f^n (y) \; = \; \int_0^y \: dy^\prime \: \hat{K} \: \otimes \:
f^{n - 1} (y^\prime)
\label{eq:a19}
\end{equation}
where the $0$-jet contribution $f^0 = \hat{f}^{(0)}$ of
(\ref{eq:a16}) and where $\hat{K}$ is the full resummed kernel of
(\ref{eq:a17}).  For the initial non-perturbative input $f^{(0)}$
in (\ref{eq:a16}) we take
\begin{equation}
f^{(0)} (y) \; = \; 3N (1 - e^{-y})^5 \: \exp (- k_T^2/Q_0^2)
\label{eq:a20}
\end{equation}
where the normalization $N$ is fixed so that the gluon,
integrated over the region $k_T^2 > Q_0^2$, carries half the
momentum of the proton.  We set $Q_0^2 = 1$ GeV$^2$.

Although the sum $f (y)$ of (\ref{eq:a18}) is independent of
$\mu$, the individual contributions $f^n (y)$ are $\mu$
dependent.  Recall that $\otimes$ stands for an integration over
$d^2 q_T$ (see (\ref{eq:a9}) and (\ref{eq:a10})), and that $f$ is
a function of $k_T^2$ as well as $y$.  In Figs.\ 3, 4 and 5 we
show the decomposition of $f (y, k_T^2)$ for $k_T = 2$, 5 and 10
GeV respectively, in each case taking three different values for
the resolution, namely $\mu = 1$, 2 and 3.5 GeV.  The gluon
density, and its decomposition, are not observable directly.  The
choices we have made for $\mu$ are, at this stage, solely to gain
insight into the structure of the BFKL gluon.  The results show
the following features:
\begin{itemize}
\item[(i)] Gluon jets with $\mu > k_T$ occur; the probability
increases as $x$ decreases.
\item[(ii)] The lower the value of $\mu$, the greater the number
of resolved jets, that is the greater the preponderance of
multijet configurations.
\item[(iii)] As $x$ decreases, the greater the diffusion in $\ln
q_T^2$ so that an $n$-jet configuration first increases in
probability and then decreases as higher jet-configurations take
over.
\item[(iv)] The higher the value of $k_T^2$ the sooner in $x$ (as
$x$ decreases) will a given multijet configuration go through
this rise and fall.
\item[(v)] As $k_T^2/\mu^2$ increases the $0$-jet contribution
drops rapidly to zero.
\end{itemize}

The results for low values of the resolution parameter $\mu$ show
that the functions $f_{n}$ have a maximum which shifts to smaller values 
of $x$ with increasing $n$. This maximum is a straightforward 
consequence of virtual corrections which, for low $\mu$, are not entirely 
compensated by (unresolved) real radiation. The maximum disappears for 
large $\mu$ and we have this structure for all values of $k_{T}$.

Some insight into the behaviour can be obtained from the analytic
form presented in section 3, which applies when $\mu^2/k_T^2$ is
small.  In this limit the virtual and unresolved real terms lead
to a suppression factor
\begin{equation}
\Delta (y) \; = \; e^{- Ay}
\label{eq:b20}
\end{equation}
where $A \equiv \overline{\alpha}_S \ln (k_T^2/\mu^2)$.  Thus
from (\ref{eq:e17}) we obtain the $0$-jet contribution
\begin{equation}
f^0 \; = \; \hat{f}^{(0)} (y, k_T^2) \; = \; e^{-Ay} \: \int_0^y
\: dy^\prime \: e^{Ay^\prime} \: 3Ne^{-k_{T}^{2}/Q_{0}^{2}} \: \frac{d}{dy^\prime}
\: (1 - e^{-y^\prime})^5 \; ,
\label{eq:c20}
\end{equation}
that is the $k_{T}$ dependence of $f^{0}$ is essentially the
same as the $k_{T}$ dependence of the driving term $f^{(0)}$ of (\ref{eq:a20}).  
This explains the origin of feature (v), that the 0-jet contribution falls 
rapidly to zero with increasing $k_{T}^{2}$. Fig.\ 6 compares the analytic 
approximation with the full result for $\mu^{2}=1$ GeV$^{2}$ and $k_{T}^{2}=4$ 
GeV$^{2}$. We see that the analytic form reproduces the shape of the numerical 
solution, but fails in the normalization.  Also the peak in the numerical 
prediction shifts slightly to smaller $x$. Thus the analytical approximation
cannot be used as a valid representation for the jet contributions, even for
a resolution as low as 1 GeV$^{2}$. 

\bigskip
\noindent {\large \bf 5.  Jet decomposition of $F_i (x, Q^2)$ at
small $x$}

We are now in a position to estimate the probability of the
different multijet configurations in the small $x$ observables
that
are driven by the BFKL gluon.  The most relevant process to study
is deep-inelastic scattering at HERA.  Using the results of
section 4, we calculate the jet decomposition of the proton
structure functions $F_i (x, Q^2)$.  In other words we determine
what fraction of events that make up the inclusive measurement of
$F_i (x, Q^2)$ contain no-jets, one jet, two jets etc.\ as a
function of $x, Q^2$ and the jet resolution parameter $\mu$. 
Recall that our jets are gluons emitted with transverse momentum
$q_T > \mu$.

From knowledge of the BFKL gluon $f$ we can determine the
behaviour of the structure functions via the $k_T$ factorization theorem, see 
Fig.\ 2.  For the transverse and longitudinal functions we have
\begin{equation}
F_{T,L} (x, Q^2) \; = \; \int_x^1 \: \frac{dx^\prime}{x^\prime}
\: \int \: \frac{dk_T^2}{k_T^4} \: f \left ( \frac{x}{x^\prime},
k_T^2 \right ) \: F_{T,L}^{\gamma g} (x^\prime, k_T^2, Q^2)
\label{eq:a21}
\end{equation}
where, to lowest order, photon-gluon fusion $F^{\gamma g}$ is
given by the quark box (and crossed box) contributions, as shown
in Fig.\ 7.  To carry out the integration over the quark line in
Fig.\ 7 we express its four momenta $\kappa$ in terms of the
Sudakov variables
$$
\kappa \; = \; \alpha p \: - \: \beta q^\prime \: + \:
\mbox{\boldmath $\kappa$}_T
$$
where $q^\prime = q + xp$ and $p$ are the basic light-like
momenta ($q$ and $p$ are the 4-momenta of the virtual photon and
proton respectively).  The variable $\alpha$ is fixed by the
quark mass-shall constraint, leaving integrations over $\beta$
and $\mbox{\boldmath $\kappa$}_T$.  Evaluating the box
contributions, equation (\ref{eq:a21}) then becomes \cite{akms}
\begin{eqnarray}
\label{eq:a22}
F_T (x, Q^2) & = & 2 \: \sum_q \: e_q^2 \: \frac{Q^2}{4 \pi} \:
\int_{k_0^2} \: \frac{dk_T^2}{k_T^4} \: \int_0^1 \: d\beta \:
\int \: d^2 \kappa_T^\prime \: \alpha_S \: f \left (
\frac{x}{x^\prime}, k_T^2 \right ) \nonumber \\
& & \nonumber \\
& \times & \left \{ \biggl [\beta^2 + (1 - \beta)^2 \biggr ] \:
\left [ \frac{\kappa_T^2}{D_1^2} \: - \: \frac{\mbox{\boldmath
$\kappa$}_T \cdot (\mbox{\boldmath $\kappa$}_T - \mbox{\boldmath
$k$}_T)}{D_1 D_2} \right ] \: + \: \frac{m_q^2}{D_1^2} \: - \:
\frac{m_q^2}{D_1 D_2} \right \} \\
& & \nonumber \\
\label{eq:a23}
F_L (x, Q^2) & = & 2 \: \sum_q \: e_q^2 \: \frac{Q^4}{4 \pi} \:
\int_{k_0^2} \: \frac{dk_T^2}{k_T^4} \: \int_0^1 \: d\beta \:
\beta^2 (1 - \beta)^2 \: \int \: d^2 \kappa_T^\prime \nonumber \\
& & \nonumber \\
& \times & \alpha_S \: f \left ( \frac{x}{x^\prime}, k_T^2 \right
) \; \left \{ \frac{1}{D_1^2} \: - \: \frac{1}{D_1 D_2} \right \}
\end{eqnarray}
where the denominators
\begin{eqnarray}
D_1 & = & \kappa_T^2 \: + \: \beta (1 - \beta) Q^2 \: + \: m_q^2
\nonumber \\
D_2 & = & \left | \mbox{\boldmath $\kappa$}_T - \mbox{\boldmath
$k$}_T \right |^2 \: + \: \beta (1 - \beta) Q^2 \: + \: m_q^2
\nonumber
\end{eqnarray}
and where $\kappa_T^\prime = \kappa_T - (1 - \beta) k_T$.  The
$x^\prime$ integration of (\ref{eq:a21}) is implicit in the $d^2
\kappa_T^\prime$ and $d\beta$ integrations.  Indeed $x^\prime$ is
fixed in terms of $\kappa_T^\prime$ and $\beta$
\begin{equation}
x^\prime \; = \; \left [ 1 \: + \: \frac{\kappa_T^{\prime 2} +
m_q^2}{\beta (1 - \beta) Q^2} \: + \: \frac{k_T^2}{Q^2} \right
]^{-1},
\label{eq:a24}
\end{equation}
which ensures that the requirement $0 < x^\prime < 1$ is
satisfied.  Of course the integration regions of (\ref{eq:a22})
and (\ref{eq:a23}) must be constrained by the condition
\begin{equation}
x^\prime (\beta, \kappa_T^{\prime 2}, k_T^2, Q^2) \; > \; x
\label{eq:a25}
\end{equation}
so that the argument $z = x/x^\prime$ of $f$ satisfies the
requirement $z < 1$.  In (\ref{eq:a22}) and (\ref{eq:a23}) we sum
over the quark flavours; we take the masses to be $m_q = 0$ for
$u,
d, s$ quarks and $m_c = 1.5$ GeV for the charm quark.  The
argument of $\alpha_S$ is taken to be $\kappa_T^{\prime 2} +
m_0^2$, which allows integration over the entire region of
$\kappa_T^{\prime 2}$.  For the light quarks we take $m_0 = 1$
GeV$^2$; the results are not very sensitive to variations of
$m_0$ about this value.  For the charm quark contribution we set
$m_0^2 = m_c^2$.  Also we set $k_0^2 = 1$ GeV$^2$.

The jet decomposition of $F_{L,T}$ are simply
obtained by substituting the $n$-jet unintegrated distribution
$f^n$ into (\ref{eq:a22}) and (\ref{eq:a23}).  In this way we can
break down the observables into their component $n$-jet
contributions, for example for $F_2 = F_L + F_T$ we have
\begin{equation}
F_2 \; = \; \sum_{n = 0}^\infty \: F_2^n.
\label{eq:b25}
\end{equation}
Figs.\ 8 and 9 show the components $F_2^n (x, Q^2)$ for deep
inelastic events containing $n$ observed jets, where in the upper
plot we require the jets to have $q_T > 3.5$ GeV, whereas in the
lower plot we demand $q_T > 6$ GeV.  Figs.\ 8 and 9 correspond to
$Q^2 = 10$ and 20 GeV$^2$ respectively.  For these choices of jet
resolution it can be seen that the $0$-jet configuration
dominates.  That is most of the emission from the BFKL ladder is
in the form of unresolved and virtual gluon radiation.  As
expected the $n$-jet configurations first become important (with
decreasing $x$) for the lower resolution, $\mu = 3.5$ GeV, and
the higher $Q^2$ value, $Q^2 = 20$ GeV$^2$, and begin to compete
with the $0$-jet rate for $x \lower .7ex\hbox{$\;\stackrel{\textstyle
<}{\sim}\;$} 10^{-5}$.  In fact the
4-jet rate becomes comparable with the $0$-jet rate for $x \sim
10^{-6}$.

Although the $0$-jet configuration dominates in the HERA
kinematic regime, there is still a non-negligible contribution
from resolved jets.  For example, at $Q^2 = 10$ GeV$^2$ and $x =
2 \times 10^{-4}$ the 1 and 2 jet contributions are each
approximately $\frac{1}{3}$ of the $0$-jet rate, and even
the 3 and 4 jet configurations occur at a reasonable rate. Also notice 
the production of resolvable jets with $\mu^{2} 
\lower .7ex\hbox{$\;\stackrel{\textstyle
>}{\sim}\;$} Q^{2}$ is important --- this is a straightforward consequence
of diffusion in $k_{T}^{2}$. 

The experiments at HERA show that the (inclusive) structure
function $F_2$ rises as $x$ decreases.  How is this rise made up
from the various multijet configurations?  First we look at the
results for the lower jet resolution, $\mu = 3.5$ GeV.  Although
the $0$-jet rate dominates, its increase with decreasing $x$ is
relatively weak compared to the data.  The rise of $F_2$ comes
from the increasing importance of the higher jet configurations. 
On the other hand at the higher resolution, $\mu = 6$ GeV, the
$0$-jet configuration is even more dominant and shows a steeper
rise over the same $x$ range, as is required for consistency of
the results.  This characteristic difference should hopefully be
seen in the measurement of the individual jet structure
functions.

The cross section for deep-inelastic scattering is readily
calculated from $F_{T,L}$.  We have
\begin{equation}
\sigma \; = \; 4 \pi \alpha^2 \: \int \: \frac{dx}{x} \: \int \:
\frac{dQ^2}{Q^4} \: \left \{ y^2 \: x \: F_1 (x, Q^2) \: + \: (1
- y) \: F_2 (x, Q^2) \right \}
\label{eq:a26}
\end{equation}
where as usual $y = Q^2/xs$, $F_T = 2x F_1$ and $F_L = F_2 - 2x
F_1$.  Here we present results for the component cross sections
$\sigma^n$ for deep-inelastic events containing $n$ jets with
$q_T > \mu$, again for two choices of resolution $\mu = 3.5$ and
6 GeV.  We take $\sqrt{s} = 300$ GeV and integrate $\sigma$ over
the interval $0.01 < y < 0.5$ so as to approximately reproduce the
HERA domain.  Figs.\ 10, 11, 12 and 13 show respectively the 0, 1,
2 and 3 jet cross sections integrated over
$x$ and $Q^2$ bins of size $\Delta x = 2 \times 10^{-4}$ and
$\Delta Q^2 = 10$ GeV$^2$, where the two entries in each bin correspond 
to a gluon jet with resolution $\mu=3.5$ and 6 GeV respectively.  We see 
that there are an appreciable number of identifiable jets.  For example, 
if we take a resolved jet to be one with $q_T > 3.5$ GeV and an integrated 
luminosity ${\cal L} = 10$ pb$^{-1}$, then in the bin defined by $0.8 \times
10^{-3} < x < 10^{-3}$ and $15 < Q^2 < 25$ GeV$^2$ we predict
1052, 790, 392 events containing 1, 2, 3 jets as compared to 6790
events with no identifiable jet.

Recall that the predictions are obtained by numerically solving the 
BFKL equation for the gluon.  The normalisation is dependent of the choice 
of the cut-off.  Here we have taken the cut-off to be 1 GeV$^2$, which was found to 
give a satisfactory description of the inclusive $F_2$ distribution.  However, 
the fraction of events containing 0, 1, 2, \ldots identifiable gluon jets is 
independent of the choice of the cut-off.  For example, for the above $(\Delta x, 
\Delta Q^2)$ bin and for the lower jet resolution of $\mu = 3.5$ GeV we find 
75\% of the cross section contains no observable jet and that 1, 2 and 3 jets 
occur 11, 8, 4\% of the time respectively.  Only 2\% of the events contain 
more than 3 jets.  For the higher jet resolution of $\mu = 6$ GeV we predict 
that the BFKL chain will give 90\% of the events with no observable jet, 
leaving only 10\% of the total to be split between 1, 2, \ldots jet events.

We see from Figs.\ 11 and 12 that the 2-jet rate is comparable to the 1-jet 
rate, and moreover that the 2-jet/1-jet ratio increases with increasing 
resolution $\mu$.  This type of behaviour is consistent with the expectations 
of the conservation of transverse momentum. \\

\bigskip
\noindent {\large \bf 6.  Summary and Conclusions}

In this paper we have formulated a modified form of the BFKL equation which 
allows an exclusive analysis of the multijet yields in deep-inelastic lepton 
scattering in the small $x$ regime.  The jets are defined as gluon emissions 
from the BFKL chain which have transverse momenta $q_T$ greater than a specified 
resolution $\mu$.  We first solved the modified BFKL equation to determine the 
jet decomposition of the unintegrated gluon distribution $f (x, k_T^2)$.  We 
then used the $k_T$-factorization theorem to determine the jet decomposition 
of the structure function $F_2 (x, Q^2)$ and of the total deep-inelastic cross 
section in the HERA small $x$ regime.  We presented the jet decompositions as a 
function of the kinematic variables and for different choices of the jet 
resolution parameter $\mu$.

The modified BFKL equation is shown symbolically in (\ref{eq:a15}) and the 
kernel $\hat{K}$ in (\ref{eq:a17}).  The equation embodies a resummation of 
the virtual contributions together with the {\it unresolved} real gluon 
emissions with $q_T < \mu$.  As a consequence the kernel $\hat{K}$ has an 
explicit $y = \ln 1/x$ dependence, which depends on the amount of unresolved 
radiation and so is a function of $\mu$.  Indeed for unrealistically low values 
of $\mu$ we derived, for pedagogic purposes, the analytic form of the $y$ 
dependence of the kernel, see (\ref{eq:c17}).  For the more realistic numerical 
solutions that we present the correlation between the $x$ dependence of the 
$n$-jet cross sections and the resolution parameter $\mu$ is apparent.

The behaviour of the $n$-jet contribution to the gluon $f$, or to $F_2$, exhibits 
a characteristic behaviour as $x$ decreases, rising to a maximum and then falling 
back to zero.  The higher the value of $n$ the lower the value of $x$ at which 
the maximum occurs.  In the HERA small $x$ regime the behaviour is only apparent 
for low choices of the parameter $\mu$, for example $\mu \sim 1$ GeV, see Figs.\ 
3, 4 and 5.  For experimentally realistic values of the resolution parameter 
(say $\mu = 3.5$ or 6 GeV) the maxima shift to very small values of $x$.  The 
dominant contribution in the HERA range then comes from events with no resolved 
gluon jets emitted from the BFKL chain.  Nevertheless the 1, 2, 3, \ldots jet 
rates are still significant.  An interesting feature of the multijet cross 
sections is that they are non-negligible even if $\mu > Q$.  The existence 
of such jets with $q_T > Q$ is a straightforward consequence of the characteristic 
$\ln k_T^2$ diffusion along the BFKL gluon chain.

To sum up, we have made an exploratory study of a form of the BFKL equation 
which allows the final state jet configurations to be determined in a consistent 
manner.  We solved the equation and presented sample results to illustrate the 
properties of these gluon jets which occur in deep inelastic scattering at 
small $x$ as a result of BFKL dynamics.  Of course there are subleading 
$\ln 1/x$ and fixed-order QCD jet contributions to consider.  These may modify 
the predictions in the HERA regime, but with decreasing $x$ the BFKL behaviour 
should become increasingly dominant.  One non-leading effect is the imposition 
of the constraint $q_{nT}^2 < x_n k_T^2/x$ (in the notation of Fig.\ 1) which 
follows from the requirement that the virtuality of the gluon links is 
dominated by $-k_T^2$ \cite{KMSGLU,LUND}.  If this were done we find that it 
would limit the available phase space for multijet production and, as a 
consequence, reduce the yield of multijet events. \\

\newpage
\noindent {\large \bf Acknowledgements}

J.K.\ thanks the Department of Physics and Grey College of the University of 
Durham, and C.A.M.L.\ thanks the H.\ Niewodniczanski Institute of Nuclear 
Physics in Krakow, for their warm hospitality, and thanks the UK Particle Physics 
and Astronomy Research Council for a Studentship.  This work has been supported 
in part by Polish KBN grant no. 2 P03B 231 08 and the EU under contracts nos. 
CHRX-CT92-0004 and CHRX-CT93-0357. \\

\vspace*{0.5cm}
\noindent {\large \bf Appendix: Numerical techniques used to
solve the BFKL equation}

Here we briefly describe the numerical method that we used to
solve the BFKL equation (\ref{eq:a15}).  The starting point is
the Chebyshev polynomial expansion of the unintegrated gluon
distribution $f (y, k_T^2)$ in which we map the region $Q_0^2 <
k_T^2 < Q_f^2$ into the interval ($-1, 1$) in terms of the
variable
$\tau$ defined by
\renewcommand{\theequation}{A1}
\begin{equation}
\tau (k_T^2) \; = \; 2 \ln \left ( \left . \frac{k_T^2}{Q_f
Q_0} \right ) \; \right / \; \ln \left ( \frac{Q_f^2}{Q_0^2}
\right ).
\label{eq:append1}
\end{equation}
We expand the gluon distribution $f$ in the polynomial form
\renewcommand{\theequation}{A2}
\begin{equation}
f (y, k_T^2) \; = \; \sum_{i = 1}^N \: C_i \biggl (\tau (k_T^2)
\biggr ) \: f_i (y)
\label{eq:append2}
\end{equation}
where $f_i (y)$ are the values of $f (y, k_T^2)$ at the
$(k_T^2)_i$ nodes obtained from
\renewcommand{\theequation}{A3}
\begin{equation}
\frac{(k_T^2)_i}{Q_f Q_0} \; = \; \left ( \frac{Q_f}{Q_0} \right
)^{\tau_i},
\label{eq:append3}
\end{equation}
with $\tau_i$ defined by
\renewcommand{\theequation}{A4}
\begin{equation}
\tau_i \; = \; \cos [ (i \: - \: {\textstyle \frac{1}{2}})
\pi/N],
\label{eq:append4}
\end{equation}
and $N$ the number of terms in the Chebyshev polynomial.  The
$k_T^2$ dependent functions $C_i$ are obtained from the Chebyshev
polynomial functions
\renewcommand{\theequation}{A5}
\begin{equation}
T_n (\tau) \; = \; \cos \biggl (n \arccos (\tau) \biggr )
\label{eq:append5}
\end{equation}
and are given by
\renewcommand{\theequation}{A6}
\begin{equation}
C_i (\tau) \; = \; \frac{2}{N} \: \sum_{n = 1}^N \: \nu_n T_n
(\tau) \: T_n (\tau_i),
\label{eq:append6}
\end{equation}
where $\nu_n = 1$ for $n > 1$, and $\nu_1 = \frac{1}{2}$.  A good
approximation for the $k_T^2$ dependence of $f$ is obtained with
typically $N = 20$.

The expansion (\ref{eq:append2}) is then substituted into the
BFKL equation (\ref{eq:a15}) to give the discretised (symbolic)
form
\renewcommand{\theequation}{A7}
\begin{equation}
f_i (y) \; = \; f_i^{(0)} (y) \: + \: \int_0^y \: dy^\prime \:
\sum_{i = 1}^N \: \hat{K}_{i,k} (y - y^\prime) \: f_k (y^\prime),
\label{eq:append7}
\end{equation}
where the full kernel (\ref{eq:a17}) now becomes
\renewcommand{\theequation}{A8}
\begin{equation}
\hat{K}_{i,k} \; = \; \sum_l \: [e^{(y - y^\prime) K_{UV}}]_{i,l}
\: K_{l,k}^R
\label{eq:append8}
\end{equation}
and the input distribution $\hat{f}_i^{(0)} (y)$ of
(\ref{eq:a16}) is
\renewcommand{\theequation}{A9}
\begin{equation}
f_i^{(0)} (y) \; = \; \int_0^y \: dy^\prime \: \sum_k \: [ e^{(y
- y^\prime) K_{UV}} ]_{i,k} \: \frac{\partial f_k^{(0)}
(y^\prime)}{\partial y^\prime}.
\label{eq:append9}
\end{equation}
The substitution of (\ref{eq:append2}) into (\ref{eq:a7}) and
(\ref{eq:a8}) gives the explicit form of the kernels $K_R$ and
$K_{UV}$ respectively.  The BFKL equation (\ref{eq:append7}) is
a Volterra-type integral equation, which we solve iteratively for
the $f_i (y)$'s.  The gluon distribution $f (y, k_T^2)$ is then
reconstructed from (\ref{eq:append2}).

We also use a Chebyshev interpolation to calculate the $Y \equiv
y - y^\prime$ dependence of the matrix elements of exponential
matrix in (\ref{eq:append8}) and (\ref{eq:append9}).  For
convenience we denote the matrix elements
\renewcommand{\theequation}{A10}
\begin{equation}
[e^{Y \: K_{UV}}]_{i,k} \; \equiv \; M (Y)_{i,k}.
\label{eq:append10}
\end{equation}
As before we expand in terms of Chebyshev polynomials
\renewcommand{\theequation}{A11}
\begin{equation}
M (Y)_{i,k} \; = \; \sum_{j = 1}^J \: C_j (\tau (Y)) \: M_{i,k}^j
\label{eq:append11}
\end{equation}
where $M^j$ are the values of $M(Y)$ at the nodes $Y_j$.  Here we
take $J = 10$.  We map the relevant region $0 < Y < Y_{\rm max}$, where
$Y_{max} = log (1/X_{min})$,
into the interval $-1 < \tau < 1$ by choosing
\renewcommand{\theequation}{A12}
\begin{equation}
\tau (Y) \; = \; (2Y \: - \: Y_{\rm max})/Y_{\rm max}.
\label{eq:append12}
\end{equation}
The $C_j$ is given by (\ref{eq:append6}) (together with
(\ref{eq:append4}) and (\ref{eq:append5})) with $i$ replaced by
$j$.  It remains to calculate $M(Y)$ at the nodes $Y = Y_j$.  We
do this by solving  
\renewcommand{\theequation}{A13}
\begin{equation}
\frac {\partial{M_{i,k} (Y)}}{\partial {y}} \; = \; \sum_{j=1}^{J}
(K_{UV})_{i,j} \: M_{j,k}(Y) .
\label{eq:append13}
\end{equation}
using the Runge-Kutta method with the boundary condition $M_{i,k}
(Y=0)=I_{i,k}$.

\newpage
\noindent {\large \bf Figure Captions}

\begin{itemize}
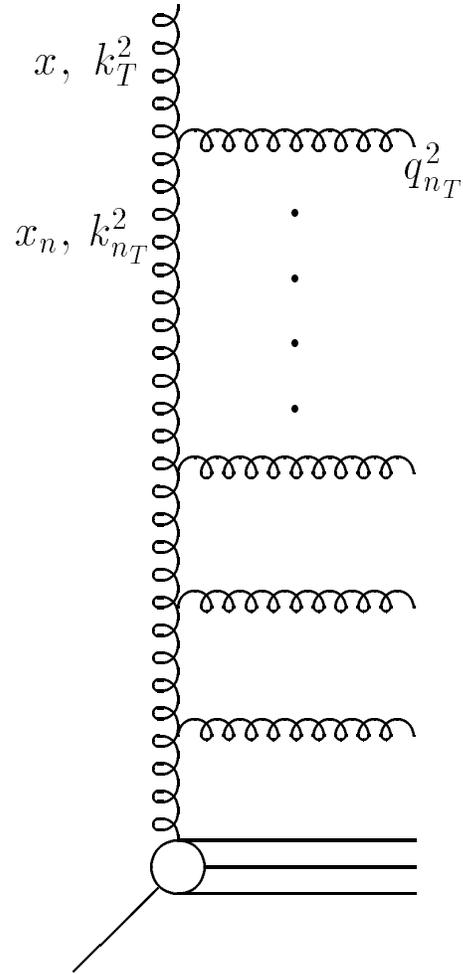
\item[Fig.\ 1] The unintegrated gluon distribution, $f (x, k_T^2)$, is effectively 
the sum of the ladder diagrams formed by the modulus squared of such amplitudes.  
The leading $\alpha_S \ln 1/x$ resummation is accomplished by the BFKL equation.

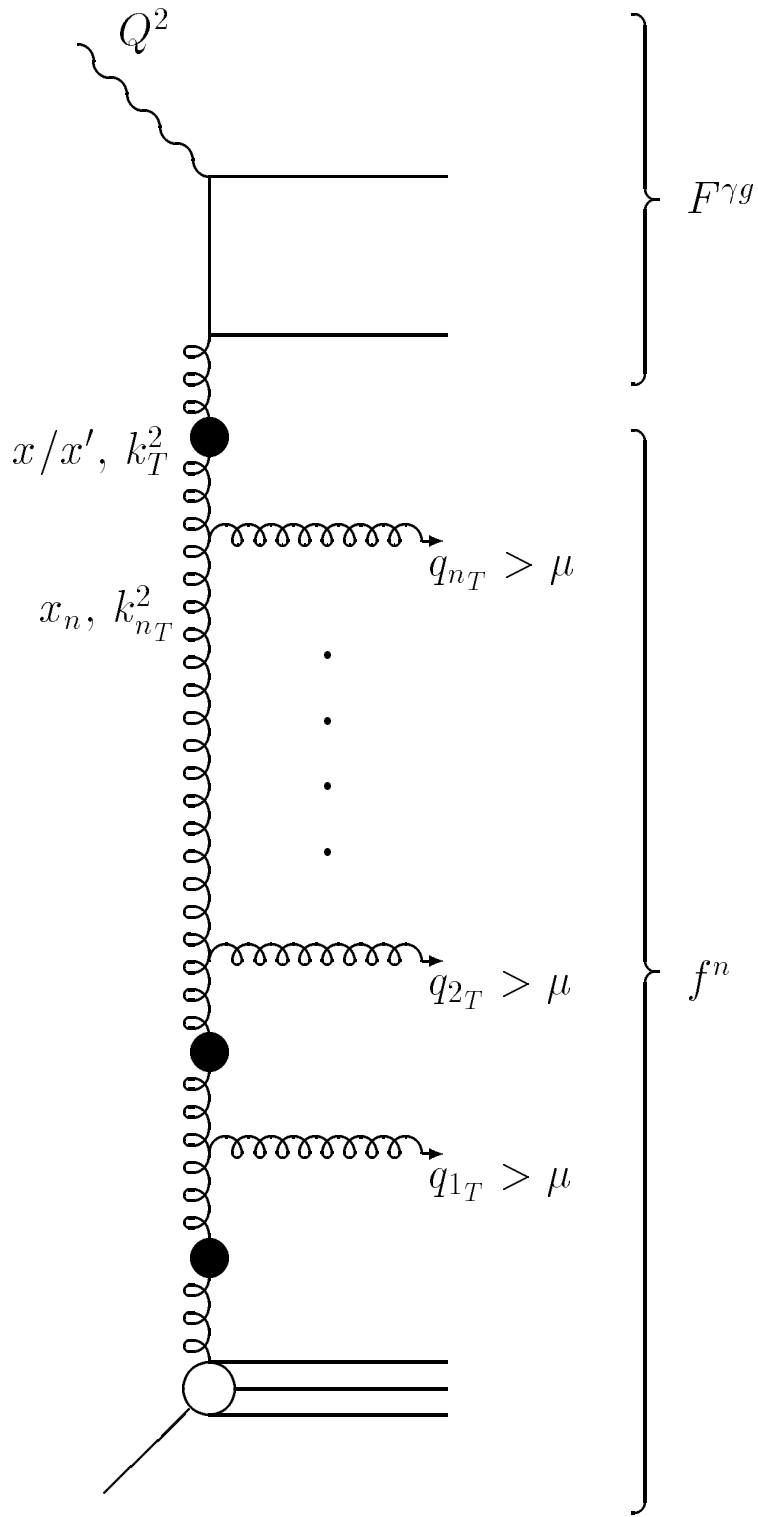
\item[Fig.\ 2] The modulus squared of this diagram gives the component $F_i^n$ 
of the proton structure function $F_i$ which arises from the contribution $f^n$ 
to the gluon distribution $f$ in which there are $n$ {\it resolved} gluon jets 
emitted along the BFKL chain, that is $n$ gluons with $q_T > \mu$.  The black 
circles are to indicate the presence of both virtual and 
unresolved gluon emissions.  The component $F_i^n$ is calculated by the 
$k_T$-factorization theorem, which has the symbolic form $F_i^n = F_i^{\gamma g} 
\otimes f^n$, see (\ref{eq:a21}) and (\ref{eq:b25}).

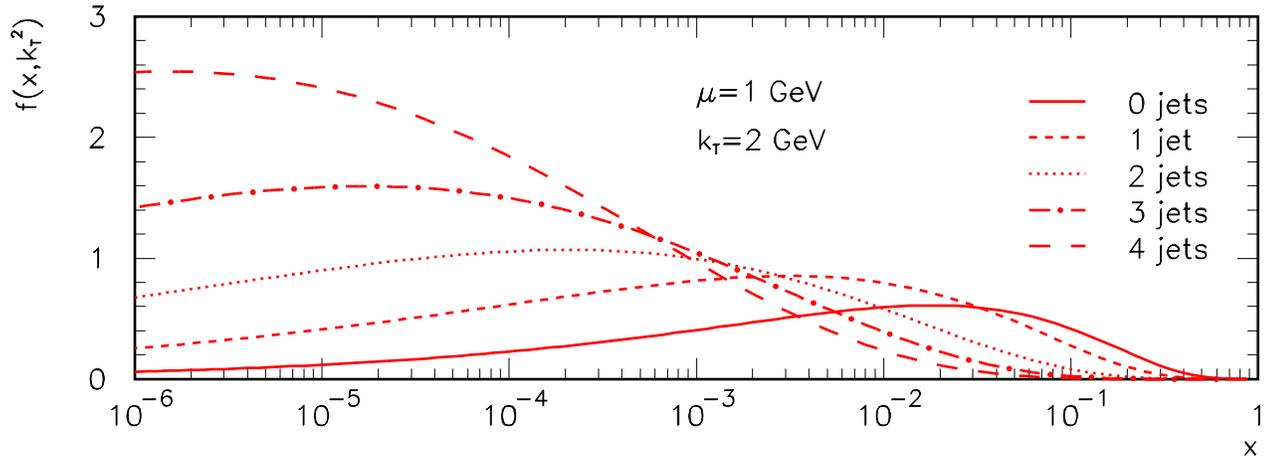
\item[Fig.\ 3] The $n$-jet contributions to the unintegrated gluon distribution 
$f (x, k_T^2)$ for three different values of the jet resolution parameter $\mu$ 
and for $k_T = 2$ GeV.

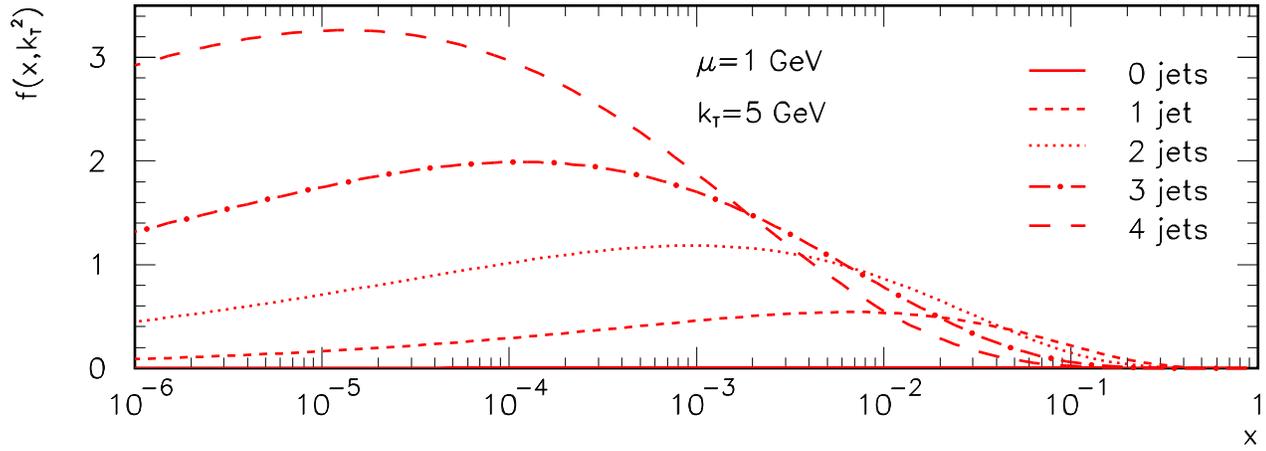
\item[Fig.\ 4] The same as Fig.\ 2 but for $k_T = 5$ GeV.

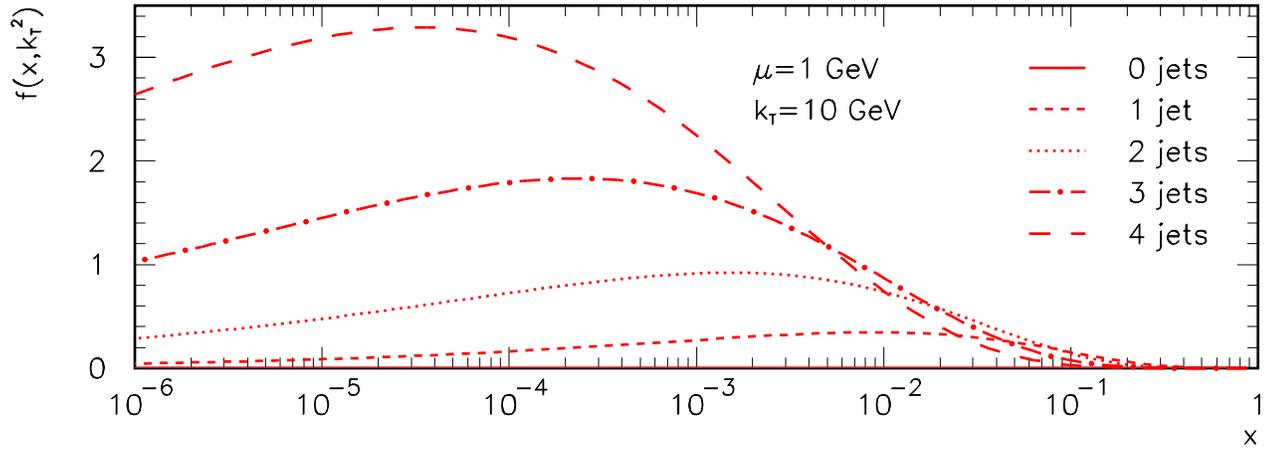
\item[Fig.\ 5] The same as Fig.\ 2 but for $k_T = 10$ GeV.

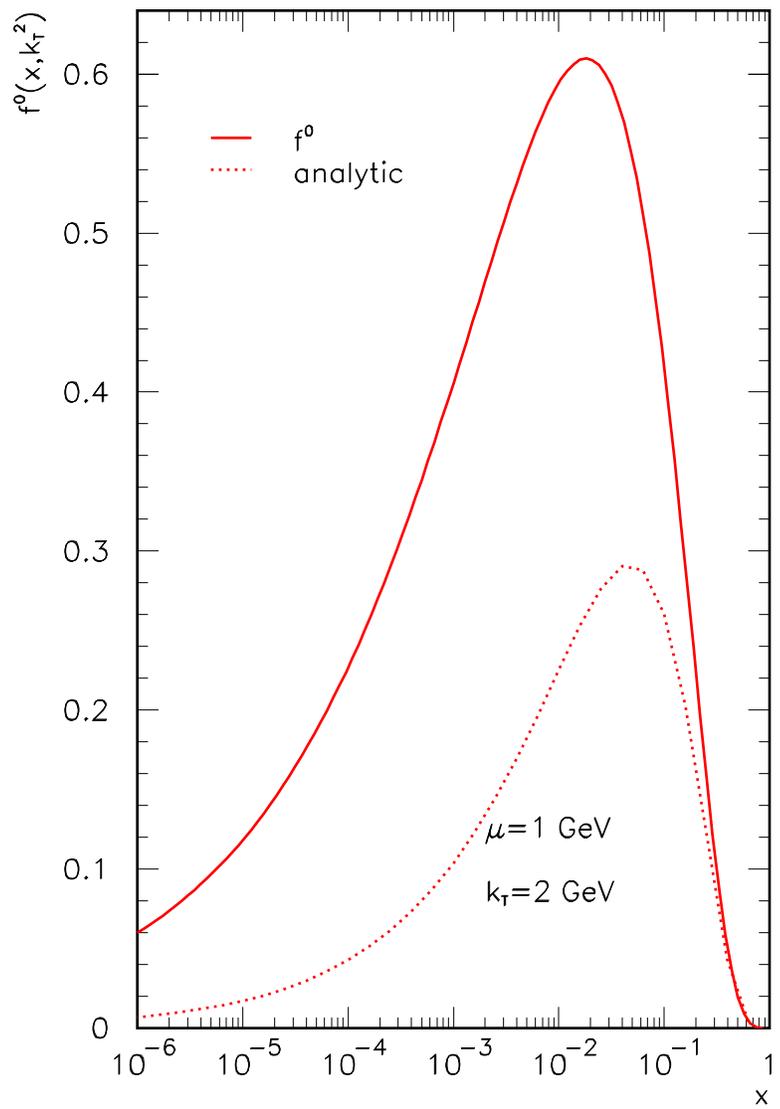
\item[Fig.\ 6] The comparison of the analytic and numerical solutions for the 
$0$-jet contribution $f^0 (x, k_T^2)$ to the unintegrated gluon distribution.

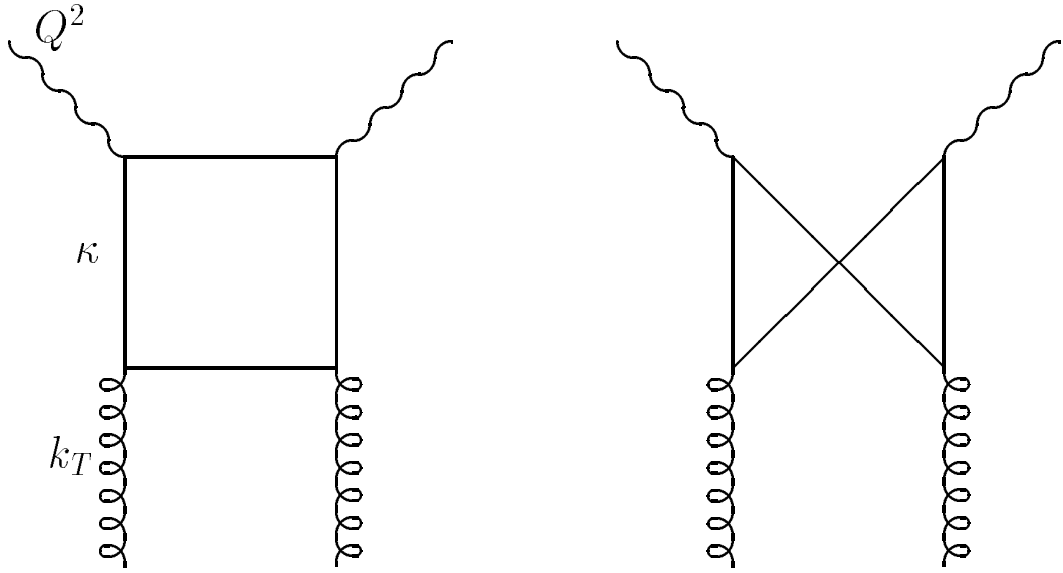
\item[Fig.\ 7] The quark box and crossed box diagrams describing photon-gluon 
fusion, $F^{\gamma g}$ in (\ref{eq:a21}).

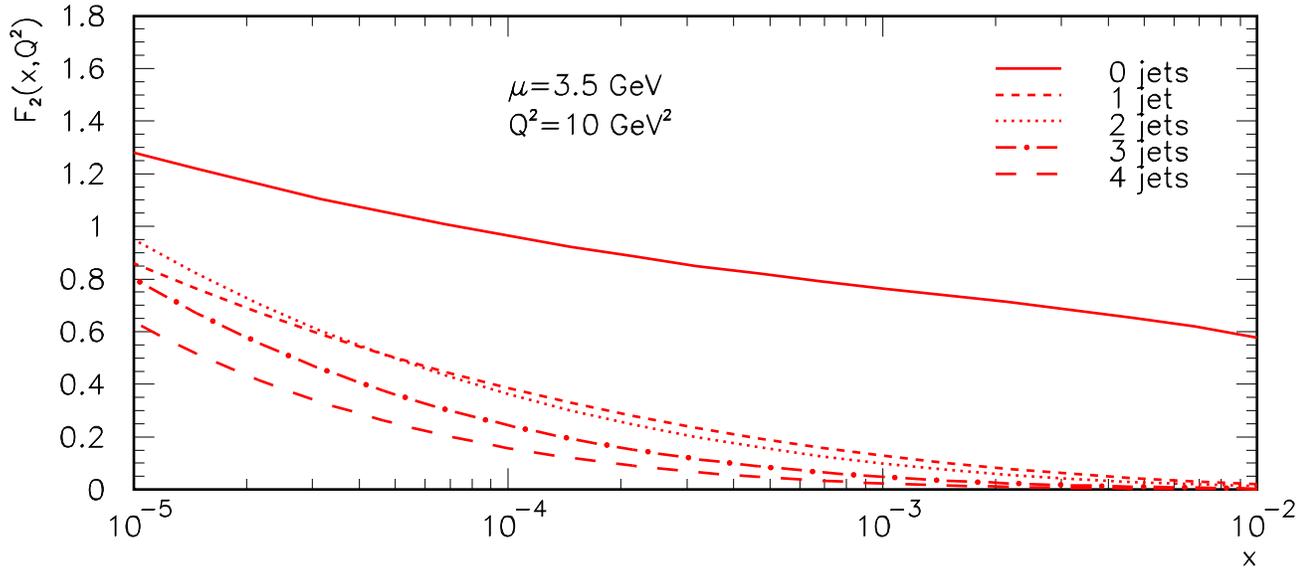
\item[Fig.\ 8] The decomposition of the proton structure function $F_2 (x, Q^2)$ 
into contributions coming from different numbers of resolved gluon jets for 
experimentally accessible values of the resolution parameter $\mu = 3.5$ and 
6 GeV.  The decomposition is shown as a function of $x$ for $Q^2 = 10$ GeV$^2$.

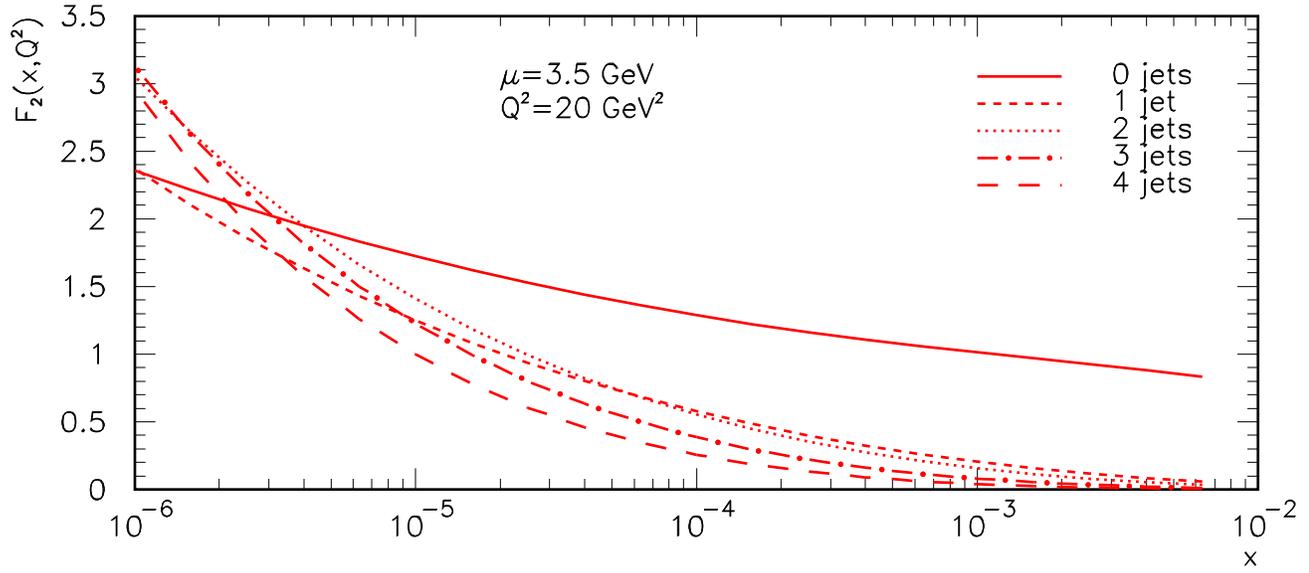
\item[Fig.\ 9] The same as Fig.\ 8 but for $Q^2 = 20$ GeV$^2$.

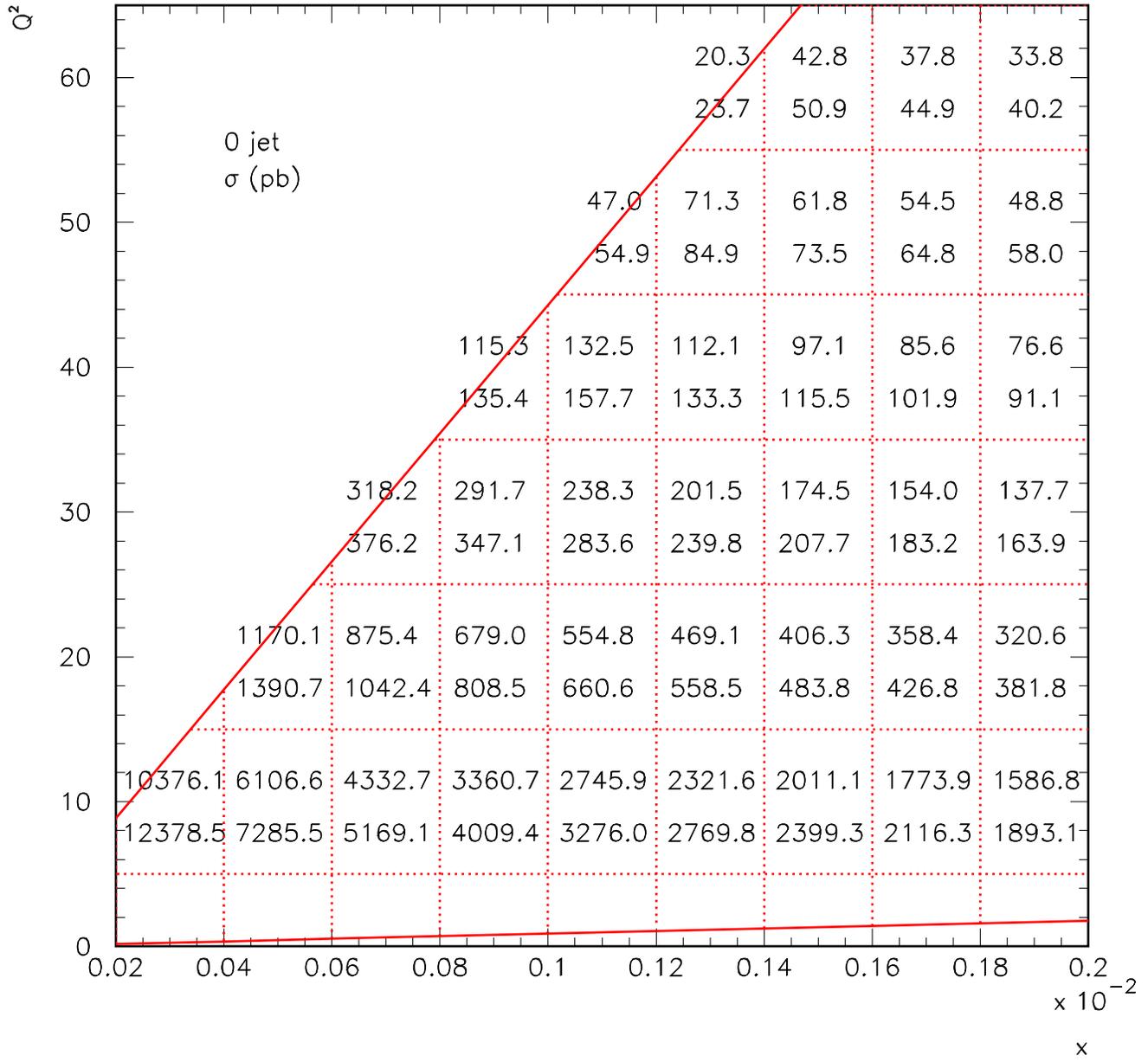
\item[Fig.\ 10] The cross-section (in pb) for deep-inelastic scattering in 
which there are no resolved gluon jets shown in different $x, Q^2$ bins in 
the region accessible at HERA.  The width of the bins are $\Delta Q^2 = 10$ 
GeV$^2$ and $\Delta x = 2 \times 10^{-4}$.  The upper and lower values correspond 
to the resolution parameter $\mu = 3.5$ and 6 GeV respectively.

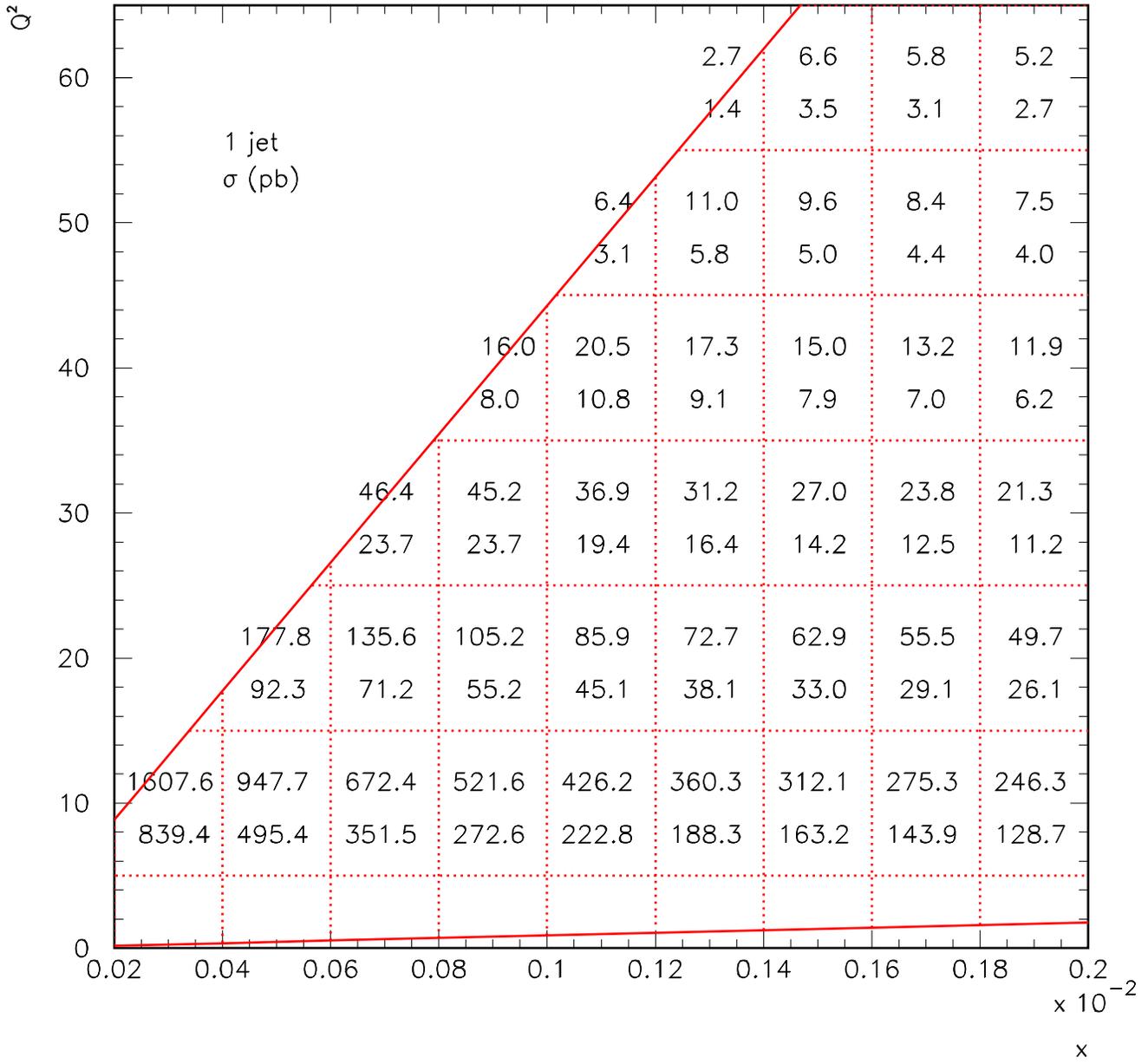
\item[Fig.\ 11] The same as Fig.\ 10 but from the contribution in which there 
is one, and only one, gluon resolved jet with $q_T > \mu$.

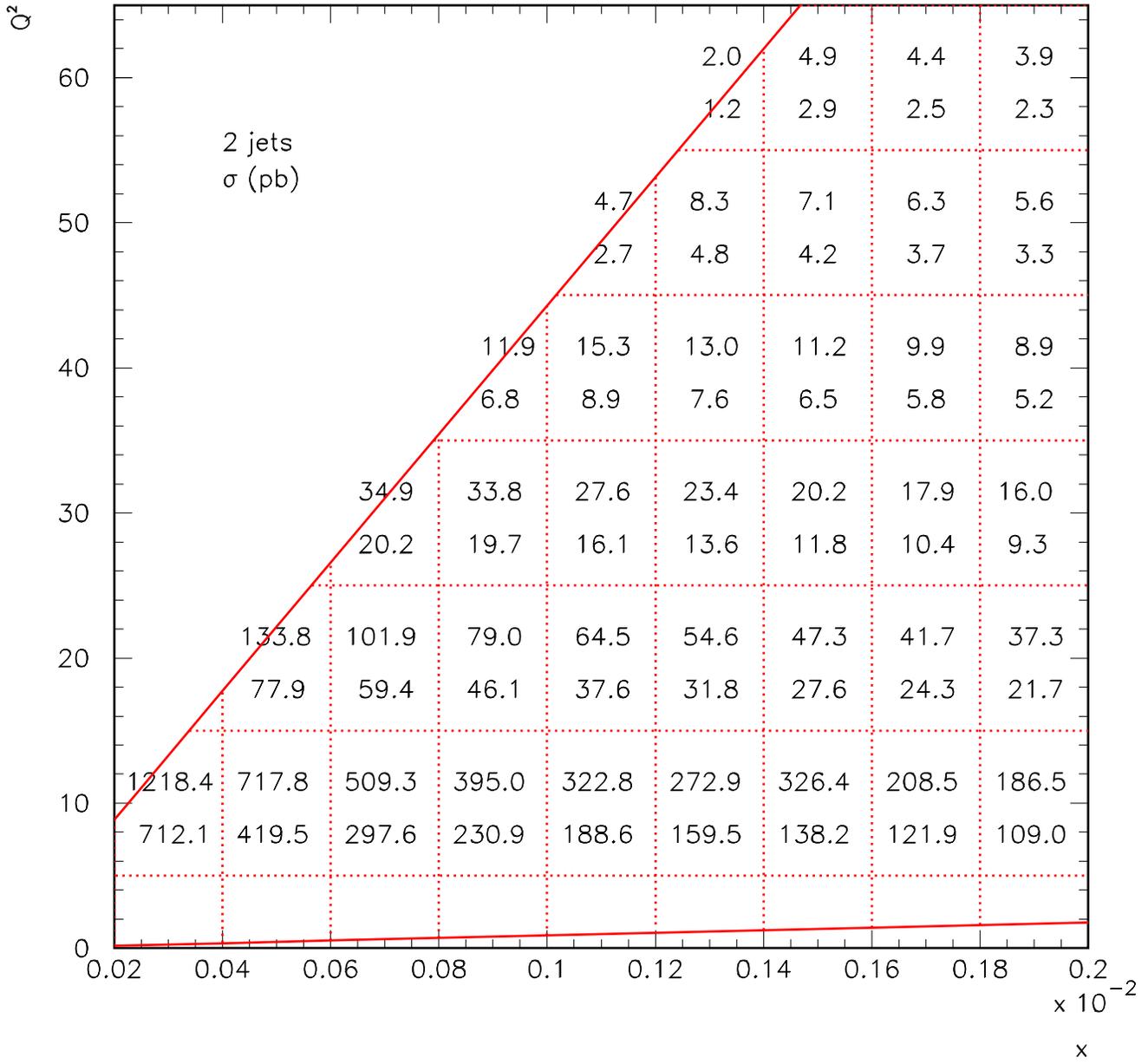
\item[Fig.\ 12] The same as Fig.\ 10 but for two resolved gluon jets.

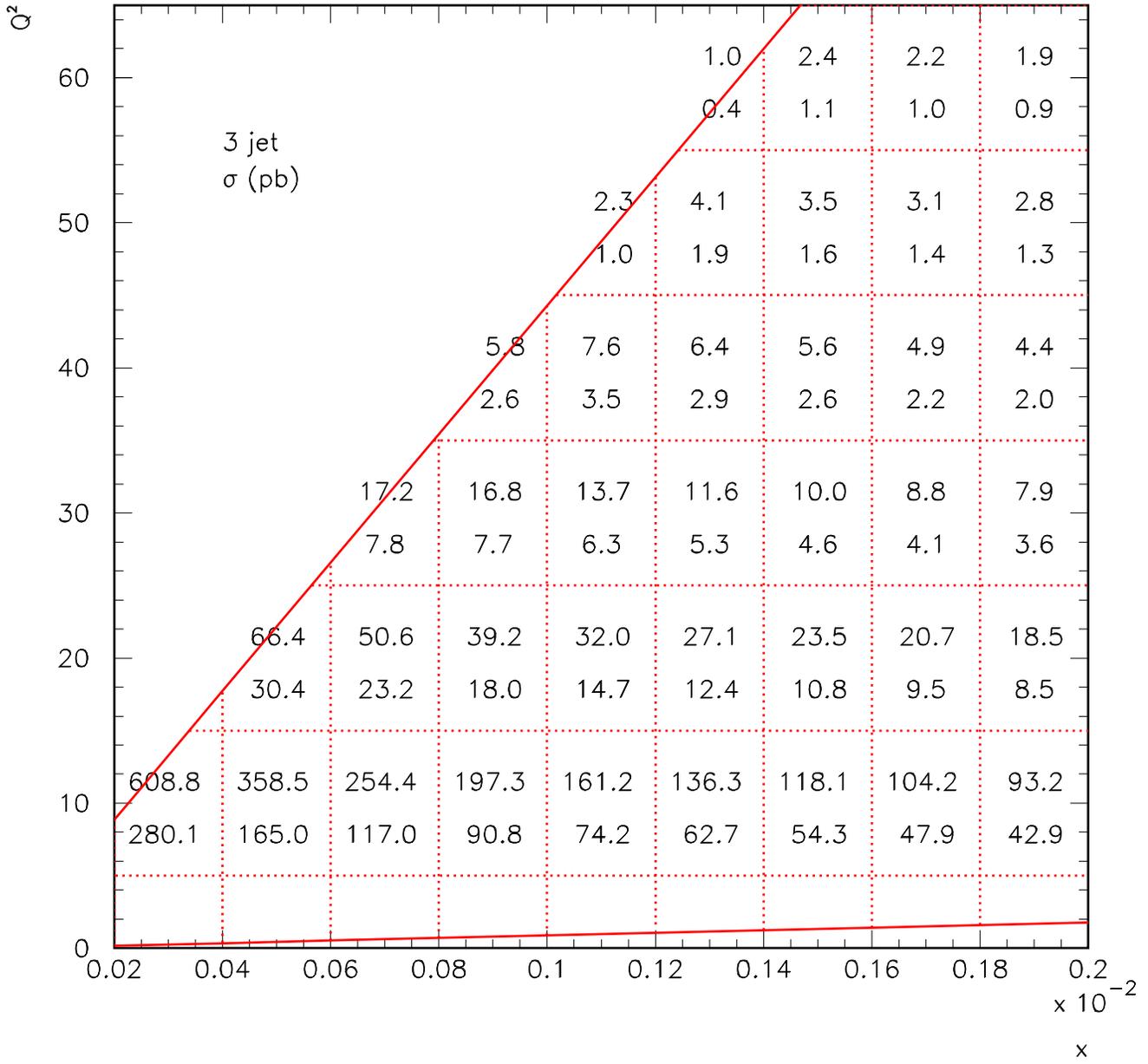
\item[Fig.\ 13] The same as Fig.\ 10 but for three resolved gluon jets.
\end{itemize}


\newpage
\begin{thebibliography}{xx}
\bibitem{h1} H1 collaboration: S.\ Aid {\it et al.}, DESY report
96-039, Z.\ Phys.\ {\bf C} (in press).

\bibitem{zeus} ZEUS collaboration: M.\ Derrick {\it et al.}, Z.\ Phys.\ 
{\bf C69} (1996) 607, DESY report 96-076, Z.\ Phys.\ {\bf C} (in press).

\bibitem{bfkl} E.A.\ Kuraev, L.N.\ Lipatov and V.S.\ Fadin, Phys.
Lett. {\bf B60} (1975) 50; Sov. Phys. JETP {\bf 44} (1976) 443;
Sov. Phys. JETP {\bf 45} (1977) 199. \\
Ya. Ya.\ Balitsky and L.N.\ Lipatov, Sov. J. Nucl. Phys. {\bf 28}
(1978) 822.

\bibitem{MRS}A.D. Martin, R.G. Roberts and W.J. Stirling, 
Phys. Rev. {\bf D50} (1994) 6734; Phys. Lett. {\bf 354} (1995) 155. 

\bibitem{GRV}M. Gl\"uck, E. Reya and A. Vogt, Z.Phys. {\bf C67} (1995) 433.

\bibitem{CCFMF2}J. Kwieci\'nski, A.D.Martin and P.J. Sutton, 
Phys. Rev. {\bf D53} (1996) 6094. 

\bibitem{KMSGLU}J. Kwieci\'nski, A.D. Martin, P.J. Sutton, Durham preprint 
DTP/96/02 (Z.Phys. {\bf C}, in print).

\bibitem{LUND}B. Andersson, G. Gustafson and J. Samuelsson, Lund preprint 
LU TP 95-13. 

\bibitem{fs} J.\ Kwieci\'nski, Nucl.\ Phys.\ {\bf B}, Proc.\
Suppl.\ {\bf 39 BC} (1995) 58; \\
M.\ Kuhlen, Proc.\ of DIS95 Workshop, Paris 1995, eds.\ J.F.\
Laporte and Y.\ Sirois, p.345.

\bibitem{ktfac} S.\ Catani, M.\ Ciafaloni and F.\ Hautmann, Phys.
Lett. {\bf B242} (1990) 97; Nucl. Phys. {\bf B366} (1991) 657; \\
J.C.\ Collins and R.K.\ Ellis, Phys. Lett. {\bf B360} (1991) 3;
\\
E.M.\ Levin, M.G.\ Ryskin and A.G.\ Shuvaev, Sov. J. Nucl. Phys.
{\bf 53} (1991) 657.

\bibitem{CIAF}M. Ciafaloni, Nucl. Phys. {\bf B296} (1988) 49. 

\bibitem{CCFM}S. Catani, F. Fiorani and G. Marchesini, 
Phys. Lett. {\bf B234} (1990) 339; Nucl. Phys. {\bf B336} 
(1990) 18;

\bibitem{JAR}T. Jaroszewicz, Acta. Phys. Polon. {\bf B11} (1980) 965. 

\bibitem{akms} A.J.\ Askew, J.\ Kwieci\'nski, A.D.\ Martin and
P.J.\ Sutton, Phys. Rev. {\bf D47} (1993) 3775.
 
\end{thebibliography}
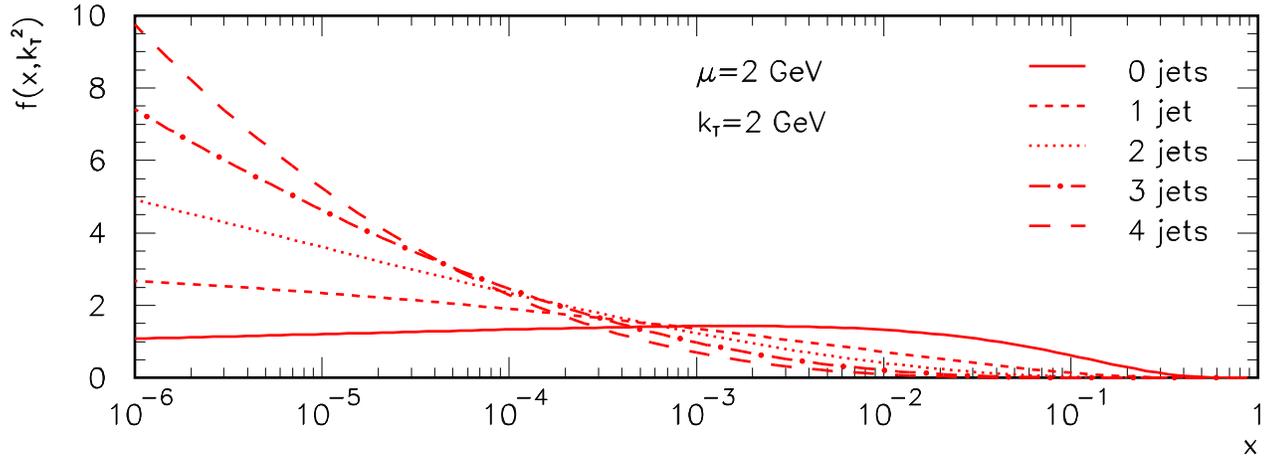
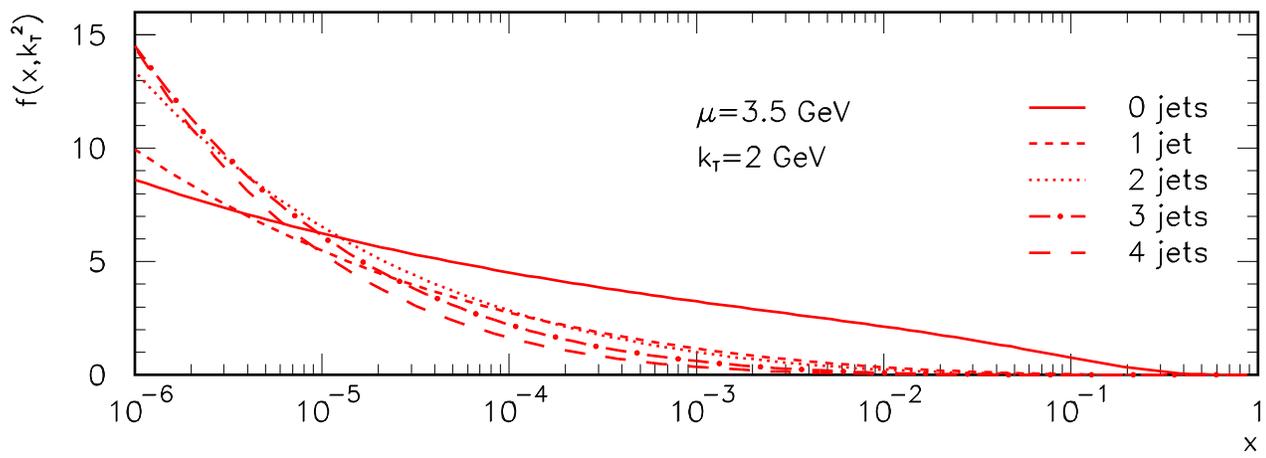
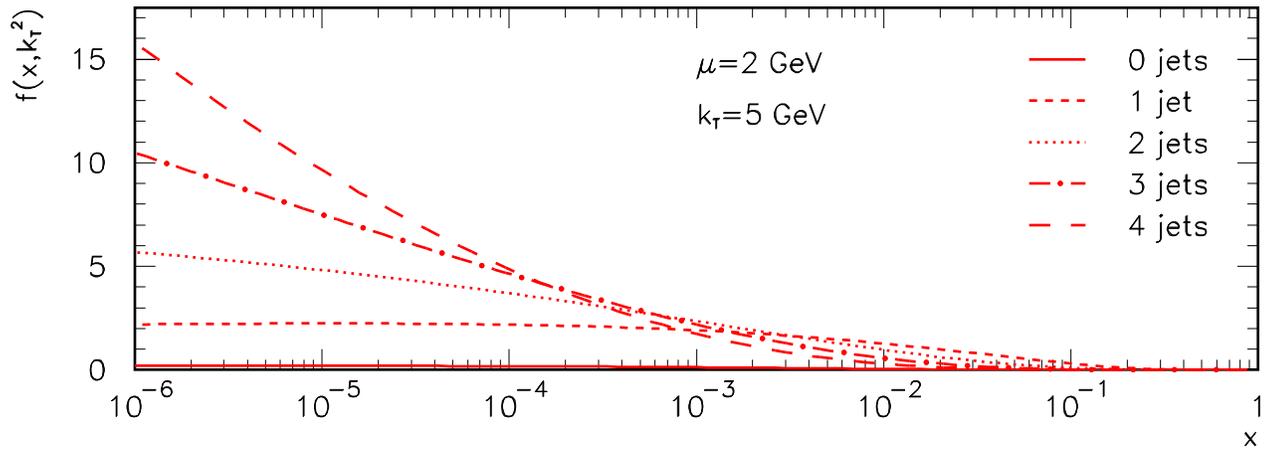
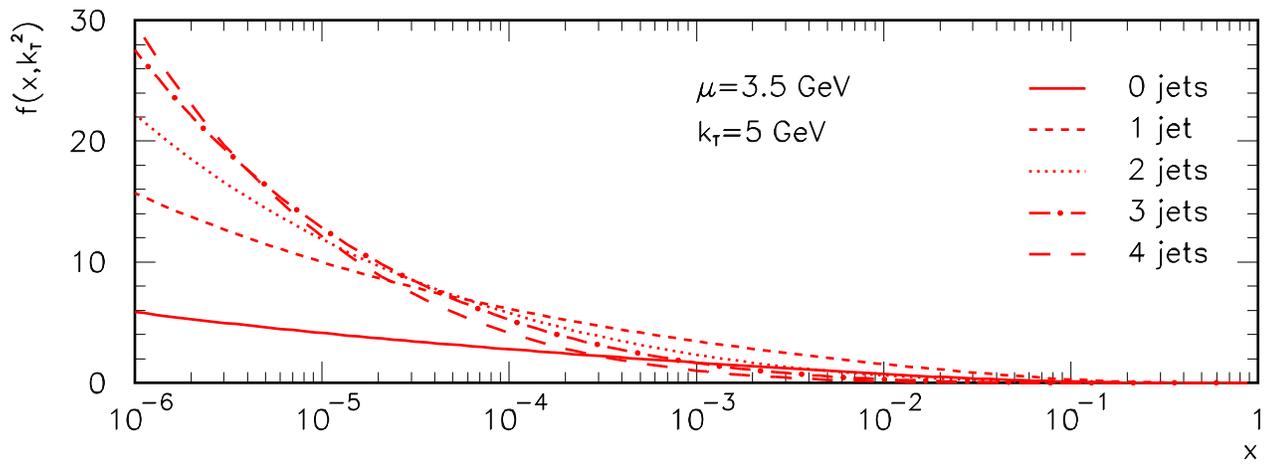
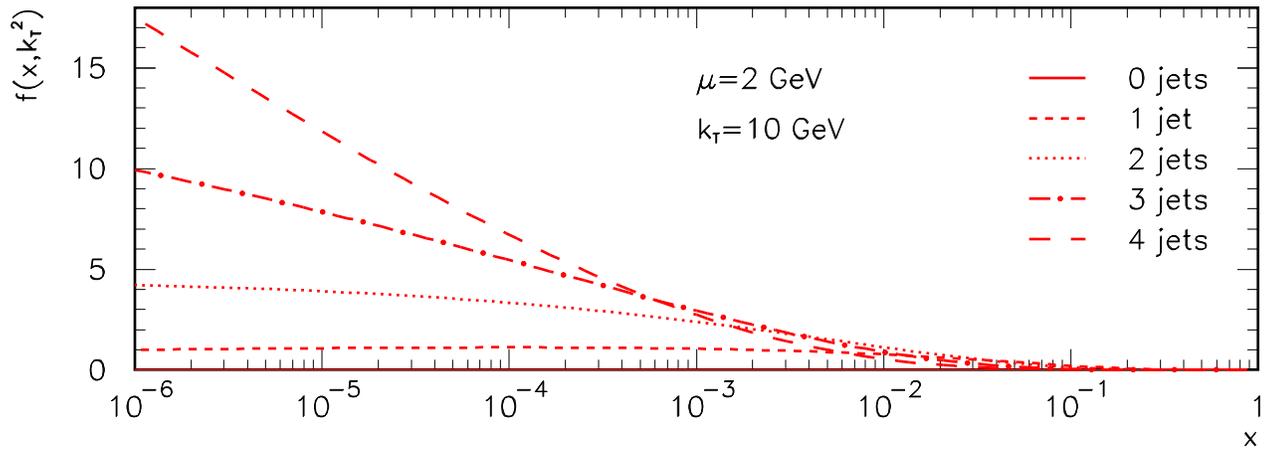
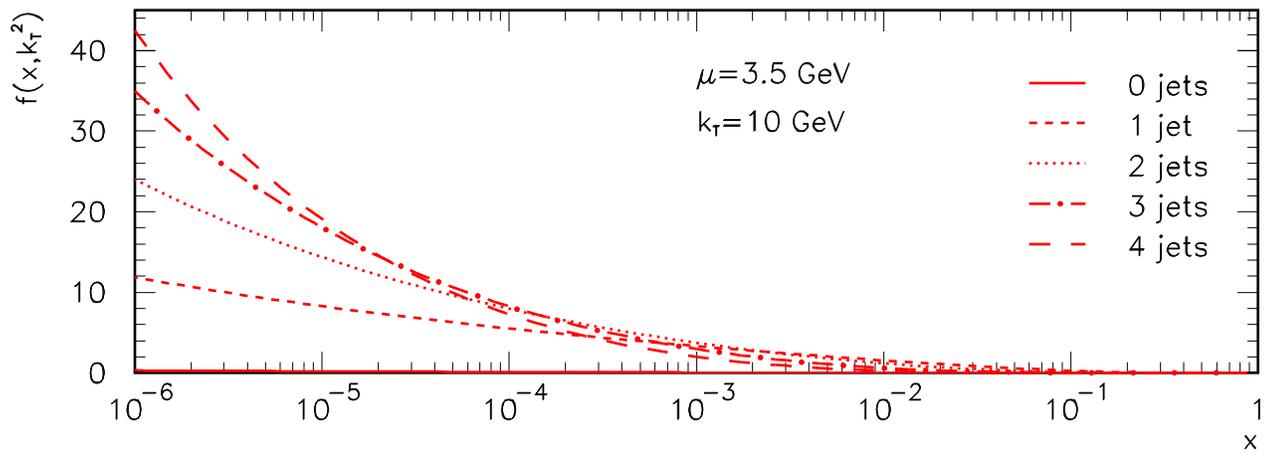
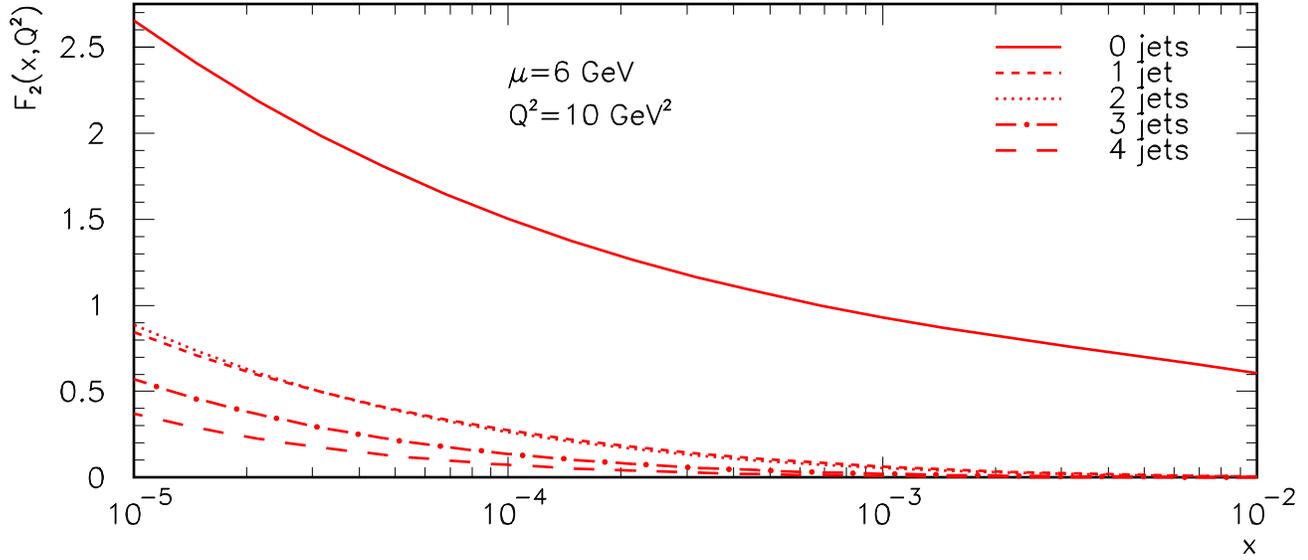
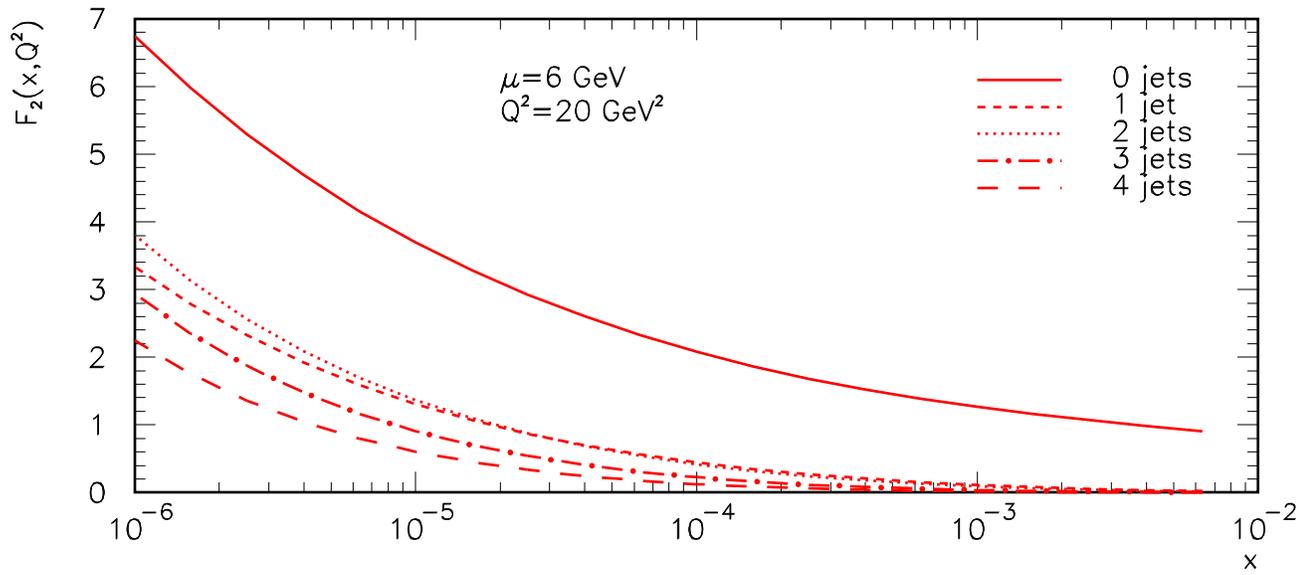
\end{document}